\documentclass[fleqn,usenatbib]{mnras}

\usepackage[T1]{fontenc}

\DeclareRobustCommand{\VAN}[3]{#2}
\let\VANthebibliography\thebibliography
\def\thebibliography{\DeclareRobustCommand{\VAN}[3]{##3}\VANthebibliography}

\usepackage{graphicx}	% Including figure files
\usepackage{amsmath}	% Advanced maths commands
\usepackage{amssymb}	% Extra maths symbols

\usepackage{color}
\usepackage[dvipsnames]{xcolor}

\newcommand{\angstrom}{\mbox{\normalfont\AA}}

\newcommand{\vunit}{\mbox{\,km\,s$^{-1}$}}

\newcommand{\Msun}{\mbox{\,$M_\odot$}}
\newcommand{\Lsun}{\mbox{\,$L_\odot$}}

\newcommand{\mic}{\mbox{$\,\mu$m}} 

\newcommand{\ebar}{\mbox{$\overline{\bf e}$}}

\usepackage{graphicx}	% Including figure files
\usepackage{amsmath}	% Advanced maths commands
\usepackage{amssymb}	% Extra maths symbols

\title[Transition radiation in novae]{Transition radiation in
dusty novae with non-thermal radio emission}

\author[A. Evans]{
A. Evans,$^{1}$\thanks{E-mail: a.evans@keele.ac.uk} \\
$^{1}$Astrophysics Group, Lennard Jones Laboratory, Keele
University, Keele, Staffordshire, ST5 5BG}

\date{Accepted XXX. Received YYY; in original form ZZZ}

\pubyear{2025}

\begin{document}
\label{firstpage}
\pagerange{\pageref{firstpage}--\pageref{lastpage}}
\maketitle

% Abstract of the paper
\begin{abstract}
Transition radiation is produced when a relativistic 
charged particle enters or leaves a solid medium.
The electrons that produce synchrotron radiation may 
interact with the dust in circumstellar environments,
leading to the emission of transition radiation.
We explore the production of transition radiation
in dusty novae that also display synchrotron radiation emission.
Transition radiation is emitted in the hard ultra-violet/X-ray
range. We suggest that, even when the transition radiation
is not itself directly observable, it may have a role in
determining the ionisation balance of, and grain heating in,
nova ejecta. Furthermore, it may be important in other dusty
environments (such as supernova remnants) with
non-thermal radio emission.
\end{abstract}


%%%%%%%%%%%%%%%%%%%%%%%%%%%%%%%%%%%%%%%%%%%%%%%%%%

%%%%%%%%%%%%%%%%% BODY OF PAPER %%%%%%%%%%%%%%%%%%

\section{Introduction}

Nova explosions occur in semi-detached binary systems 
containing a white dwarf (WD) and a cool secondary
\citep[see, e.g.,][]{CN2}. In a classical nova (CN),
the secondary is usually a dwarf, while in a recurrent
nova (RN), the secondary may be a {sub-giant} or a red giant (RG).
In each case the secondary fills its Roche lobe, and material
flows from the secondary onto the surface of the WD via
an accretion disc. Material accumulates on the WD surface,
and eventually, conditions at the base of the accreted layer
become suitable to trigger a thermonuclear runaway. The
accreted material, together with some material dredged up
from the WD, is ejected explosively in a nova eruption.
When the eruption has subsided, mass transfer resumes, 
and another explosion occurs. In CNe, the inter-eruption
time is $\sim10^4$~years, whereas in RNe, the inter-eruption
time is $\la100$~years.

Dust formation in CNe has long been known 
\citep[see, e.g.,][for reviews]{evans12,evans12a,gehrz12,evans25a}. 
The grain material can be a mixture of types
\citep{evans12,evans25a,evans12a,gehrz12}, with carbonaceous dust
being a prominent component. 
\cite{chong25} estimate that as many as $50-70$\% of CNe 
produce dust. These authors also find that dust-formation in
CNe is more likely in those novae that are $\gamma$-ray
emitters, suggesting a connection between the shocks that
result in $\gamma$-ray emission and the formation of dust.
The importance of shocks in CN dust formation has been 
stressed by \cite*{derdzinski17}. 

RNe are not, as a rule,
dust formers \citep[although there was transient dust formation
in the 2014 eruption of V745~Sco;][]{banerjee23a}. However some
RNe with RG secondaries have prominent silicate emission
features at 9.7\mic\ and 18\mic\ 
\citep[][Evans et al., to be submitted]{evans07,woodward08}.

Similarly, radio emission from both CNe and RNe has been
observed, both during quiescence and in eruption
\citep*[see, e.g.,][]{bode12,chomiuk21b}. Radio emission
during eruption may be thermal or non-thermal. The latter
is commonly synchrotron radiation, emitted by relativistic
electrons accelerated in shocks within the ejected
material.

The combination of dust and relativisic electrons 
provides an environment in which 
transition\footnote{``Transition'' in this context is
of course unconnected with the usual meaning of
transition during nova eruptions \citep[see][]{gehrz98}.}
radiation (TR) may be emitted. TR in an astrophysical
context has been discussed by (amongst others)
\cite{johansson71}, \cite{lerche72}, \cite{ramaty72}, 
\cite*{yodh73}, \cite{watson73}, and 
\cite{gurzadyan73,gurzadyan74}.

Here we explore the possible importance of TR in novae. 

\section{Transition radiation}

TR is produced when a charged particle 
crosses the boundary between two media having different
dielectic permeabilities, for example, an electron moving
into, or out of, a solid medium
\citep[see][for details]{jackson99}.
It is routinely used in transition radiation detectors
to detect and identify high energy particles 
in laboratory environments 
\citep[see, e.g.,][for a review]{andronic12}.

We begin by summarising some of the essential features
of TR. As both \cite{jackson99} and \cite{pacholczyk70} 
(see below) use cgs units, we do so here.

\begin{figure}
 \includegraphics[width=8.5cm]{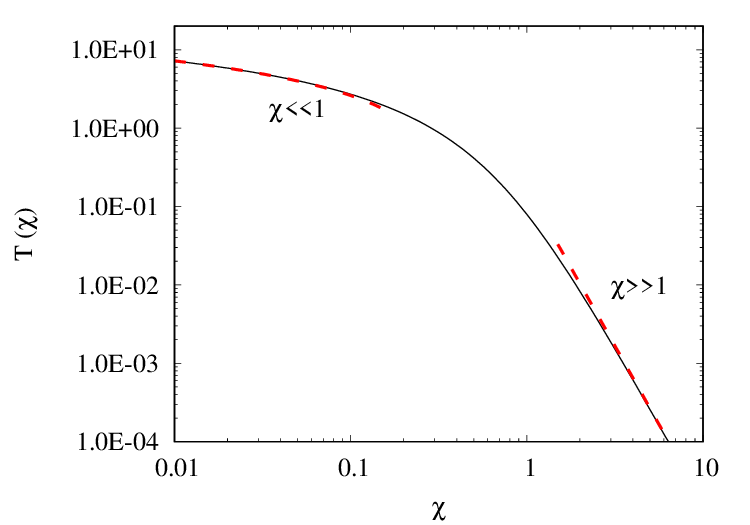}
 \caption{Spectral energy distribution of TR as a function
 of $\chi=\omega/[\gamma\omega_p]$. The two broken red curves
 are for the limiting cases $\chi\ll1$ and $\chi\gg1$.
See text for details.
 \label{TR2}}
\end{figure}

For a relativisic electron with energy $E=\gamma{mc^2}$, the
energy emitted as a function of the circular frequency
$\omega=2\pi\nu$ is \citep{jackson99}
\begin{eqnarray}
 \frac{dI}{d\omega} & = &  \frac{2e^2}{c} \left \{
 \left [ 1+2\left ( \frac{\omega}{\gamma\omega_{\rm p}} \right )^2\right ]
 \ln \left [ 1 + \left ( \frac{\gamma\omega_{\rm p}}{\omega} \right )^2 \right ] - 2 \right \}  \nonumber \\
 & = & \frac{2e^2}{c} \:\: T(\chi)  \label{TR-I}
\end{eqnarray}
where $\omega_{\rm p}$ is the plasma frequency for the solid,
and $\chi=\omega/[\gamma\omega_{\rm p}]$. 
The limiting forms of $T(\chi)$ are
\begin{eqnarray}
T(\chi) & \simeq & -2\:\ln(\ebar\chi) \mbox{~~~}(\chi\ll1) 
 \label{TR4a} \\
     & \simeq & 1/[6\chi^4] \mbox{~~~~~~~}(\chi\gg1) \:\:, \label{TR4b}
\end{eqnarray}
where \ebar{} is the base of natural logarithms.
The high frequency approximation given by 
Equation~(\ref{TR4b}) is accurate to better
than 5\% for $\chi\ga4.5$, and better than 1\% 
for $\chi\ga10$. The function given in 
Equation~(\ref{TR-I})
is plotted in Fig.~\ref{TR2}, together with the 
limiting cases given in Equations~(\ref{TR4a}) 
and~(\ref{TR4b}). Note that $T(\chi)$ diverges for
$\chi\ll1$, a consequence of the approximations made
in its derivation \citep[see][]{jackson99}. We do not 
use the $\chi\ll1$ approximation here.

The TR spectrum extends as far as, 
but not much beyond, $\chi\simeq1$, i.e., the
frequency $\omega\simeq\gamma\omega_{\rm p}$.
Since $\omega_{\rm p}\sim10^{16}$~s$^{-1}$ 
for plausible circumstellar grain materials
\citep[e.g., $\omega_{\rm p}=1.87\times10^{16}$~s$^{-1}$
for graphite;][]{duley98}, it is apparent that
the emitted radiation can potentially extend into
the ultra-violet (UV) and X-ray regions.

\begin{figure}
 \includegraphics[width=8.5cm]{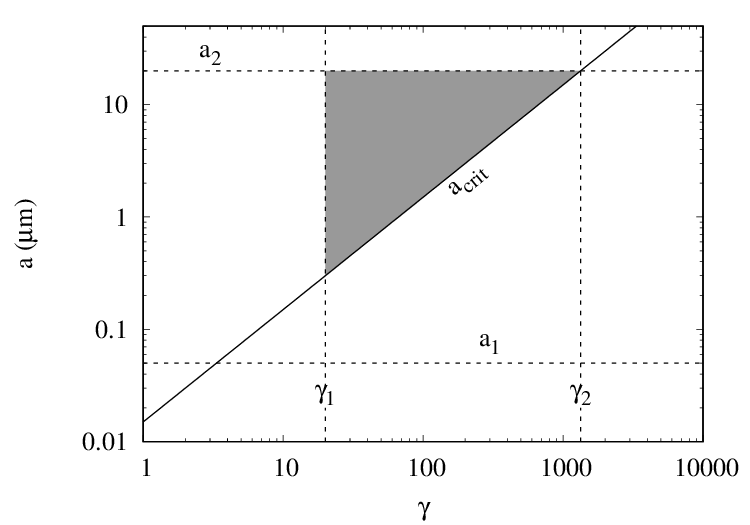}
 \caption{Schematic of the region of application of TR.
 $a_{1}$ and $a_{2}$ are respectively the minimum
 and maximum grain radius in the grain size distribution. 
 The diagonal line denotes the requirement that the grain
 radius be greater than the critical value for TR formation,
 $a_{\rm crit}$. $\gamma_{1}$, $\gamma_{2}$ denote the lowest 
 electron energy, and the highest electon energy capable of
 producing TR, respectively. 
 The parameter space for the production of TR
 is denoted by the shaded region. See text for details.\label{FZ}}
\end{figure}

TR is produced only within the ``formation zone'' of extent
\begin{equation}
 \Lambda = \gamma{c}/\omega_{\rm p} ~~.
\end{equation}
Thus an electron penetrating a spherical particle will 
produce TR only if the particle radius $a$ 
satisfies $2a>\Lambda$, or 
$a>a_{\rm crit}=c\gamma/2\omega_{\rm p}$.
Numerically
\begin{equation}
 \left ( \frac{a_{\rm crit}}{\mic} \right ) \simeq 0.015\gamma
 \left ( \frac{\omega_{\rm p}}{10^{16}\mbox{~s}^{-1}} 
 \right )^{-1} \:\:.
 \label{acrit}
\end{equation}
Even for a particle with a radius that exceeds this
critical value, TR is produced only for trajectories through
the particle that exceed $\Lambda$, so the effective grain 
cross-section for the production of TR is
\begin{equation}
 \sigma(a) = \pi{a}^2 \:\: \left [ 1 - 
 \left (  \frac{\Lambda}{2a} \right )^2 \right ] ~~~~~ 
 (a>\Lambda/2).
 \label{x-sect}
\end{equation}
This means that there are constraints on the grain 
size/electron energy parameter space in which TR
can be produced (see Fig.~\ref{FZ}). Not only is
there a lower limit on grain size, there is also 
an upper limit on the electron energy, given by
\begin{equation}
 \gamma_{\rm 2}mc^2 = \frac{2\omega_{\rm p}a_{2}}{c} \:\: mc^2 \:\:
 \simeq 340 \left ( \frac{a_2}{10\mic}\right) 
  \left ( \frac{\omega_{\rm p}}{10^{16}\mbox{~s}^{-1}} \right 
  )~\mbox{MeV~,}
 \label{g2}
\end{equation}
where $a_2$ is the largest grain radius in the grain population.

Furthermore, since an electron penetrating a grain
experiences two changes of medium (i.e., ingress into, and 
egress from, the grain), emitting TR at each interface
\citep{gurzadyan73}, there is an additional factor of two
in Equation~(\ref{TR-I}). 
The total energy emitted per unit frequency interval is therefore
\begin{equation}
\frac{dI}{d\omega}  =  \frac{4e^2}{c} \:\: T(\chi)  
\label{TRI2} \:\:.
\end{equation}

\section{Application to novae}
\subsection{Rationale}\label{rationale}

\begin{table}
 \caption{Novae with dust and
 synchrotron radiation emission.\label{dust-sr}}
\begin{tabular}{lcl}
Nova       & Year of & Reference to dust \\ 
           & Eruption & formation \\ \hline 
\multicolumn{3}{c}{\underline{Classical novae}}    \\   
V1370 Aql  & 1982 & \cite{bode84} \\
           &  & \cite{gehrz84} \\
V838 Her   & 1991 & \cite{woodward92}  \\ 
V959 Mon$^*$   & 2012 & \cite{evans24} \\ 
V1324 Sco  & 2012 & \cite{munari15} \\
        &   & \cite{finzell18} \\ 
V809 Cep   & 2013 & \cite{babul22} \\ 
V339 Del   & 2013 & \cite{gehrz15} \\
           &  & \cite{evans17} \\
V5668 Sgr  & 2015 & \cite{banerjee16} \\
           &  & \cite{gehrz18} \\
V357 Mus   & 2018 & \cite{chomiuk21a}\\ \hline
\multicolumn{3}{c}{\underline{Recurrent novae$^\dag$}} \\  
&&\\
RS Oph    & 1985, 2006, 2021 & \cite{rushton22}\\
V745 Sco  & 1989, 2014 & Evans et al., to be submitted  \\
V3890 Sgr & 1992, 2019 & \cite{kaminsky22} \\ \hline
\multicolumn{3}{l}{$^*$The dust emission in V959~Mon
was not due to dust formation,} \\
\multicolumn{3}{l}{but dust was clearly present. 
See \cite{evans24} for details.}\\
\multicolumn{3}{l}{$^\dag$The years of eruption
listed are 
only for those eruptions when} \\
\multicolumn{3}{l}{non-thermal radio emission would
have been detected. There will} \\
\multicolumn{3}{l}{have been earlier eruptions. The 
references to ``dust formation''} \\
\multicolumn{3}{l}{are to works that report the 
presence of dust in the RN} \\
\multicolumn{3}{l}{environment.}
\end{tabular}
\end{table}

Both the RNe RS~Oph and V3890~Sgr displayed synchrotron
radiation, with an implied magnetic field $B$ of a few tens
of mG. The electron gyroradius 
\begin{equation}
r_B\sim1.7\times10^5\:\gamma\:\left ( 
\frac{B}{\mbox{10mG}} \right )^{-1} \mbox{~cm,} 
\end{equation}
which is negligible by comparison with the dimensions
of the ejecta ($\sim10^{15}$~cm), even for the most
energetic electrons (although, as shown above,
the most energetic electrons do not contribute to the TR).

RS~Oph is known to possess a silicate dust shell
\citep[see, e.g.,][and references therein]{rushton22}
that seems to survive the RN eruption. The inner radius of
the RS~Oph silicate dust shell is $\sim5\times10^{14}$~cm
\citep{rushton22}, corresponding to $\sim20$~mas at a 
distance of 1.6~kpc. This is comparable with the extent
of the synchrotron-emitting
material a few tens of days after the 2006 eruption, as 
reported by \cite*{sokoloski08}. The evidence for dust in
V3890~Sgr is rather less direct, but the presence of dust
at a temperature of 400~K is required to fit the spectral
energy distribution out to $\sim30$\mic\ \citep{kaminsky22}.

Non-thermal radio emission is also seen in CNe, in which 
shocks between fast- and slow-moving flows in material
ejected in the eruption lead to particle acceleration, 
and consequently the production of synchrotron radiation
\citep{weston16}. 

\cite{chomiuk21a} take brightness temperature
$>5\times10^4$~K in erupting novae
as evidence of synchrotron radiation. The novae
(all CNe) in \citeauthor{chomiuk21a}'s compilation,
and which showed evidence of non-thermal radio emission
together with the presence of dust,
are given in Table~\ref{dust-sr}. This table also
includes the RNe that have displayed synchrotron
radiation emission and the presence of dust.

In both V1324~Sco and V357~Mus (see Table~\ref{dust-sr} for 
references), the peak of the non-thermal synchrotron 
radiation emission coincided with the deep dust minimum
in their visual light curves, suggesting a common underlying
cause for both phenomena.

The ``helium'' CN V455~Pup was extremely dusty
\citep{woudt09,banerjee23b}, the dust being located in
a bipolar shell \citep{woudt09}. V445~Pup was also a
strong radio emitter, the radio emission being dominated
for several years by synchrotron radiation \citep{nyamai21}.
\citeauthor{nyamai21} found that the synchrotron-emitting
material spatially coincided with the dusty bipolar lobes.

Of course, not every nova is likely to have temporally
and spatially coincident dust and synchrotron emission.
Even in those that do, the overlap between the dust- and
relativisic electon-bearing volumes may not be extensive,
or the grains may be too small (i.e. $a<a_{\rm crit}$;
{see Equation~(\ref{acrit}))}. However there
do appear to be a number of dusty novae with concurrent 
non-thermal radio emission to merit an exploration
of the importance that TR may have in novae.
We therefore estimate the TR produced as
the relativistic electrons responsible for the synchrotron
radio emission penetrate dust
grains in nova environments. 

\subsection{Detail}
\label{detail}
We consider a dust shell containing $N_g(a)\,da$ grains 
per unit volume with radius $a$ in the range 
$a\rightarrow{a+da}$. We assume that the dust
grains have a size distribution given by
\begin{equation}
 N_{\rm g}(a)\,da   = N_0 \, a^{-\beta}\,da ~~~ (a_{1}<a<a_{2})
 \label{size}
\end{equation}
per unit volume, where the grain radius $a$ lies between
$a_{1}$ and $a_{2}$. The exponent $\beta$ was
$\simeq2.3$ in the CN V705~Cas \citep{evans05}. From
Equation~(\ref{acrit}), the critical radius is 
$\sim1.5-15$\mic\ for $100<\gamma<1000$; grains having these
dimensions are not uncommon in CNe 
\citep[see, e.g.,][]{helton10,evans17,gehrz18}.
If the dust occupies a volume $V_{\rm g}$, the total dust
mass is
\begin{eqnarray}
 M_{\rm g} & = & \frac{4\pi\rho}{3} \:\: 
 \frac{N_0V_{\rm g}}{(4-\beta)}
\:\:\left (a_2^{(4-\beta)}-a_1^{(4-\beta)} \right ) \nonumber \\
 & \simeq & \frac{4\pi\rho}{3} \:\: 
 \frac{N_0V_{\rm g}}{(4-\beta)} \:\: a_2^{(4-\beta)}
 \label{gmass}
\end{eqnarray}
for $a_2\gg{a}_1$ and $\beta<4$. 

The maximum electron energy that will
produce TR (see Equation~(\ref{g2})) is given by 
\begin{equation}
 \gamma_2  \simeq  670 ~~ \left (  \frac{\omega_{\rm p}}
 {10^{16}\mbox{~s}^{-1}} 
 \right ) 
 \left ( \frac{a_2}{10\mic}\right )~~~.
 \label{g2a}
\end{equation}
The relativisic electron energy distribution is assumed
to be of the form
\begin{equation}
 n_e(E)\,dE = n_0 E^{-\delta} \,dE 
\end{equation}
per unit volume. If the electrons occupy a volume $V_e$,
in a source at distance $D$, the flux density in the form
of optically thin synchrotron emission is 
\begin{equation}
 [f_\nu]_{\rm SR}  =  \frac{V_e\phi}{D^2} \: 
 c_5(\delta)  n_0 B_\perp^{(\alpha+1)} \left ( \frac{\nu}{2c_1}
 \right )^{-\alpha} \label{SR}
\end{equation}
\citep[see][which includes definitions of the constants
$c_5(\delta)$ and $c_1$]{pacholczyk70}. $B_\perp$ is the
component of the magnetic field perpendicular to the direction
of the electrons' motion.
The  spectral index of optically thin synchrotron radiation
(defined as $[f_\nu]_{\rm SR}\propto\nu^{-\alpha}$) is 
$\alpha=(\delta-1)/2$.
The parameter $\phi$ is a volume filling factor 
\citep[see][for details]{chevalier98}.

The frequency of collisions between of electrons of energy
$\gamma{m}c^2$ ($\gamma\gg1$) with grains in the radius 
range $a\rightarrow{a}+da$ is $N_g(a)\sigma(a)c$.
The luminosity in the form of TR is obtained by integrating 
Equation~(\ref{TRI2}) over $a$ and $E$:
\begin{equation}
 [L_\nu]_{\rm TR}  = 8\pi{e}^2V_{\rm com} \:
 \int_E\int_a T(\chi) N_{\rm g}(a) \sigma(a) n_e(E) \:
 da\:dE \:, \label{TR3}
 \end{equation}
where an additional factor $2\pi$ has been included to
convert from 
frequency interval $d\omega$ to frequency interval $d\nu$.
The volume $V_{\rm com}$ in Equation~(\ref{TR3})
is the volume common to both the dust shell and
the synchrotron emitting region; $V_{\rm com}\le{V}_e,V_{\rm g}$.
As discussed in Section~\ref{rationale}, the observations
suggest that there could be considerable overlap
between dusty and synchrotron-emitting volumes.
Combining Equations~(\ref{SR}) and (\ref{TR3}) enables
us to scale the observed radio synchrotron radiation flux
density to estimate the expected TR.

Integrating Equation~(\ref{TR3}) using the full form
of $T(\chi)$ in Equation~(\ref{TR-I}) is cumbersome.
However as we are interested in estimating
the high energy end of the TR spectrum, we use the 
$\chi\gg1$ form of $T(\chi)$ to give
\begin{equation}
  [L_\nu]_{\rm TR}  = \frac{4\pi{e}^2V_{\rm com}}{3} \:
 \int_E\int_a 
 \left ( \frac{\gamma\omega_{\rm p}}{\omega}
 \right )^{4} N_{\rm g}(a) \sigma(a) n_e(E) \,
 dE \, da \, , \label{TR4}
\end{equation}
provided we are mindful of the constraints on $\omega$ and
$\gamma$ imposed by the condition $\chi\gg1$.
As we have noted above,
the approximation we have used for $T(\chi)$ is reasonably
good for $\chi\ga4.5$, so the application of 
Equation~(\ref{TR4}) should be a reasonable approximation
for wavelengths
\begin{equation}
\lambda \la 420 \left ( 
\frac{\omega_p}{10^{16}\mbox{~s$^{-1}$}}  \right )  \mbox{\AA} .
\label{lam-limit}
\end{equation}
The TR is emitted in the hard UV/X-ray region.

\begin{figure}
 \includegraphics[width=8.5cm,keepaspectratio]{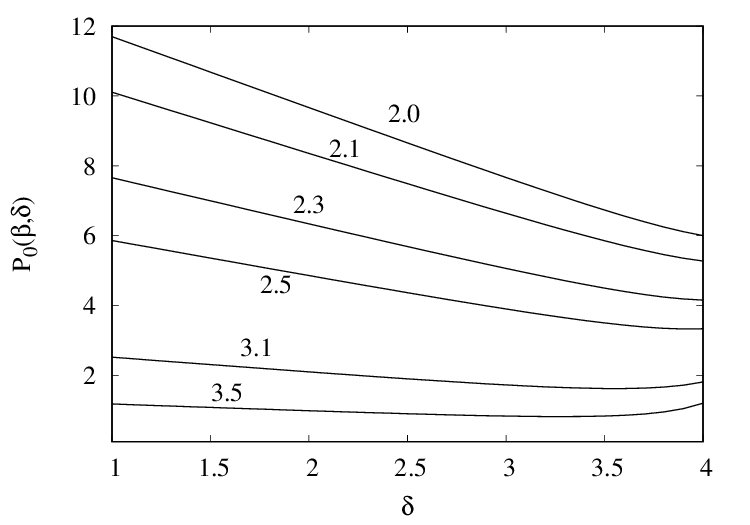}
 \caption{The function $P_0(\beta,\delta)$ 
 as a function of $\delta$
 for representative values of $\beta$. See text.\label{PP}}
\end{figure}

The integration limits in Equation~(\ref{TR4}) are
$c\gamma/[2\omega_{\rm P}]<a<a_2$, and
$E_1<E<E_2=[2\omega_{\rm p}a_2/c]\times{m}c^2$, where $E_1$ is
the lowest energy in the electron energy distribution and 
$E_2$ is the highest electron energy capable of generating
TR (see above). The region of integration is shown in 
Fig.~\ref{FZ}. For simplicity, we 
take\footnote{Justification for
this assumption is given in the Appendix.}
$E_1=0$, which gives
\begin{eqnarray}
 [L_\nu]_{\rm TR} & = & P_0(\beta,\delta) \:\: \frac{\pi{e}^2n_0}{[mc^2]^4} \: \frac{M_{\rm g}}{\rho} \: \frac{V_{\rm com}}{V_{\rm g}} \: a_2^{(4-\delta)} 
 \left ( \frac{\omega_{\rm p}}{\omega} \right )^4 \times \:\: 
 \nonumber \\
 &&   ( {2\omega_{\rm p}mc}  )^{(5-\delta)} \:\:\:.
 \label{TR5}
\end{eqnarray}
We have used Equation~(\ref{gmass}) to convert
$N_0$ to $M_{\rm g}$. The function
\begin{equation}
 P_0(\beta,\delta) =\frac{(4-\beta)}{(3-\beta)(1-\beta)}
 \left [ \frac{1-\beta}{5-\delta} - \frac{3-\beta}{7-\delta}
 + \frac{2}{8-\beta-\delta}\right ] 
 \label{P0}
\end{equation}
is plotted in Fig.~\ref{PP} for a range of $\beta$ and 
$\delta$ values. Clearly $P_0$ diverges, and may
even be negative, for some combinations of $\beta$ and $\delta$,
but the values in Fig.~\ref{PP} are typical of nova dust 
($\beta=2.0-2.5$) and of interstellar dust ($\beta=3.1-3.5$),
while the range of $\delta$ values gives radio spectral indices
for optically thin synchrotron radiation in the range $0<\alpha<1.5$.
For example, in the 2006 eruption of RS~Oph,
the radio spectral index was in the range
 $0.5\la\alpha\la0.75$ \citep*{rupen08},
 corresponding  to $2.0\la{\delta}\la2.5$.
We find that, for a reasonable range of values for
$\beta$ and $\delta$, $P_0(\beta,\delta)$ is in the range 
$1.0\la{P_0}\la12$. 

We use Equation~(\ref{TR5}) to estimate
the contribution of TR to the UV/X-ray emission of novae.

\subsection{How important is TR in novae?}
\label{estimate}
\subsubsection{Observability of TR}
To estimate the contribution of TR, 
we take $\beta=2.5$, $\delta=2.0$, 
for which $P_0=4.857$. For $\delta=2$, 
$c_5(\delta)=1.37\times10^{-23}$;
$c_1=6.27\times10^{18}$ 
\citep[both in cgs units; see Table~7 of][]{pacholczyk70}. 
Thus Equation~(\ref{TR5}) becomes
\begin{equation}
 [L_\nu]_{\rm TR}  =  4.857 \:\: \frac{\pi{e}^2n_0}{[mc^2]^4} \: \frac{M_{\rm g}}{\rho}  \: \frac{V_{\rm com}}{V_{\rm g}} \:\: a_2^{2}  \:\:  
    ( {2\omega_{\rm p}mc}  )^{3} \:\:  
   \left ( \frac{\omega_{\rm p}}{\omega} \right )^4 \:\:.
 \label{TR-nova}
\end{equation}

For the non-thermal radio emission, we take a representative 
flux density of $[f_\nu]_{\rm SR}=10$~mJy at 1~GHz for a nova
at a distance of 1~kpc, and a magnetic field of 10~mG. From
Equation~(\ref{SR}), 
\begin{equation}
 n_0V_e\phi = 6.208\times10^{38} \: \:.
 \label{nvphi}
\end{equation}
Thus (with $[L_\nu]_{\rm TR}$ in erg~s$^{-1}$~Hz$^{-1}$
and all other quantities in cgs units)
\begin{equation}
 [L_\nu]_{\rm TR}  =  2.261\times10^{39} \:\: 
 \frac{{e}^2}{[mc^2]^4} \: 
 \frac{M_{\rm g}}{\rho{r}_{\rm g}^3}  \: 
 \frac{V_{\rm com}}{V_{e}\phi} \:\: a_2^{2}  \:\:  
    ( {2\omega_{\rm p}mc}  )^{3} \:\:  
   \left ( \frac{\omega_{\rm p}}{\omega} \right )^4 \:\:,
 \label{TR-nova-2}
\end{equation}
where $r_{\rm g}$ is the radius of the dust shell, 
assumed spherical.

For the dust, we assume carbon (density $\rho=2$~g~cm$^{-3}$,
$\omega_p=1.87\times10^{16}$~s$^{-1}$) and a total
dust mass $M_{\rm g} = 50\times10^{-8}$\Msun\
\citep[see, e.g., Table~1 of][]{evans25a}. For the present,
we assume $V_{\rm com}=V_e$, also $\phi=1$
(see Section~\ref{rationale} above).
Note that, as a consequence of Equation~(\ref{nvphi}),
$[L_\nu]_{\rm TR}$
in Equation~(\ref{TR-nova-2}) is $\propto\phi^{-1}$,
so this choice of $\phi$ is the most pessimistic.

For the dust shell, we take a radius 
$r_{\rm g}=4.3\times10^{14}$~cm (the distance attained by
nova ejecta travelling at 500\vunit\ in 100~days). Thus
\begin{eqnarray}
 [L_\nu]_{\rm TR} &  = & 7.62\times10^{21} 
  \left ( \frac{r_{\rm g}}{4.3\times10^{14}~\mbox{cm}} \right )^{-3} 
  \left ( \frac{a_2}{10\mic}\right )^2 
  \nonumber \\
  && \left ( \frac{\omega_{\rm p}}{\omega} \right )^4 
   \mbox{~~erg~s$^{-1}$~Hz$^{-1}$}\:\:.
 \label{TR-nova-1}
\end{eqnarray}
For a dusty nova at 1~kpc, the photon flux in the form of TR is
\begin{eqnarray}
 [{\cal N}_\lambda]_{\rm TR} & \simeq & 9.6 \:\: 
 \left ( \frac{r_{\rm g}}{4.3\times10^{14}~\mbox{cm}} \right )^{-3} 
 \left ( \frac{a_2}{10\mic}\right )^2 
 \times \nonumber\\
 && 
 \left ( \frac{\omega_{\rm p}}{\omega} \right )^3 
  \mbox{~~photons~s$^{-1}$~cm$^{-2}$~\AA$^{-1}$.}
  \label{f-mJy1}
\end{eqnarray}
This corresponds to 
$[{\cal N}_\lambda]_{\rm TR}\simeq0.25$~photons ksec$^{-1}$ 
cm$^{-2}$~\AA$^{-1}$ at 30\AA. This flux will of course
be attenuated by the intervening interstellar and 
circumstellar columns.

The nova parameters on which $[{\cal N}_\lambda]_{\rm TR}$
depends may of course differ considerably from the values
we have used for this estimate.
Some parameters are not easily quanti\-fiable, in
particular $V_e$ and $V_{\rm com}$, while the 
effect of the parameter $\delta$ is not 
straight-forward to generalise. This is because 
(a)~$\delta$ appears as an exponent (twice) in
Equation~(\ref{TR5}), (b)~its dependency is also
determined by $c_5(\delta)$, and by $P_0$.
But given the range of possible values of
$M_{\rm g}$ ($\sim0.04\times$ to $\sim25\times$ the value
we have used here), $a_2$ (lower by a factor $\sim10$,
and noting that the TR flux depends as $a_2^{(4-\delta)}$),
$P_0$ ($\sim0.5\times$ to $\sim2\times$),
$f_\nu$ ($\sim0.1\times$ to $\sim10\times$), and
$r_{\rm g}$ ($\sim0.5\times$ to $\sim2\times$, and
noting that the TR flux depends as $r_{\rm g}^{-3}$),
we consider that the estimate in
Equation~(\ref{f-mJy1}) may range by up to three
orders of magnitude either way. We therefore recast
Equation~(\ref{f-mJy1}) in the form
\begin{equation}
 [{\cal N}_\lambda]_{\rm TR} \simeq 9.4\times10^{-6} \:\:
  K \:\: \left ( \frac{\lambda}{1\angstrom} \right )^3
   \mbox{~~photons~ksec$^{-1}$~cm$^{-2}$~\AA$^{-1}$~~,}
  \label{f-mJy2}
\end{equation}
where the parameter $K$ embodies all the nova variables.
For the numerical values we have used above, $K\equiv1$,
and the possible range of the various parameters
means that $K$ has a corresponding range of 
$\sim\pm3$~dex. It is therefore 
plausible that, in some novae, these quantities
occupy a region of parameter space that render TR
important and observable. 

Even if the estimate in Equation~(\ref{f-mJy1})
is just beyond the capabilities of current X-ray 
facilities (see, e.g., Figure~3 of 
\cite{drake21}, and Figure~5 of \cite{ness22}), TR 
should be within reach with the next generation
\citep[e.g., NewAthena;][]{cruise24}.

\subsubsection{Effect on ejecta evolution}
We also consider the importance of TR for the evolution
of the ejecta by considering how the estimated TR flux
compares with other sources of UV/X-ray emission in an
erupting nova, such as that emitted during the supersoft
phase of a nova eruption. This phase persists while nuclear
burning on the WD continues, and lasts for a time that
depends on the WD mass \citep{wolf13}; the supersoft
phase may last for tens to hundreds of days. 

For a nova, the spectral luminosity of the stellar remnant
having bolometric luminosity $L_{\rm bol}$ and effective
temperature
$T_*$ is
\begin{equation}
L_{\nu}  =  \pi{B}_\nu \: \frac{L_{\rm bol}}{\sigma{T_*}^4}\:\:;
\end{equation}
here $\sigma$ is the Stefan-Boltzmann constant. 
We take the bolometric luminosity of the stellar remnant
to be the Eddington luminosity of a 
1\Msun\ WD ($3.3\times10^4$\Lsun). For remnant effective 
temperature given by $kT_*=15$~eV ($kT_*=20$~eV),
$L_\nu\simeq1.5\times10^{14}$~erg~s$^{-1}$~Hz$^{-1}$
($\simeq3.6\times10^{16}$~erg~s$^{-1}$~Hz$^{-1}$), both
at 30\AA. From Equation~(\ref{TR-nova-1}), the 
spectral luminosity of TR is
\begin{equation}
 [L_\nu]_{\rm TR} \simeq 7.62\times10^{21} \left
 ( \frac{\omega_p}{\omega} \right )^4
 \mbox{~~erg~s$^{-1}$~Hz$^{-1}$ ~~,}
\end{equation}
taking the other bracketed parameters to be of order unity. Thus
$[L_\nu]_{\rm TR}\simeq6\times10^{15}$~erg~s$^{-1}$~Hz$^{-1}$
at 30\AA\ (and noting the considerable range possible
in the nova parameters that determine the importance of TR). 
It seems that TR in a 
dusty nova environment may in some cases be comparable
with that of radiation from the stellar remnant.

\begin{figure}
 \centering
 \includegraphics[width=8cm,keepaspectratio]{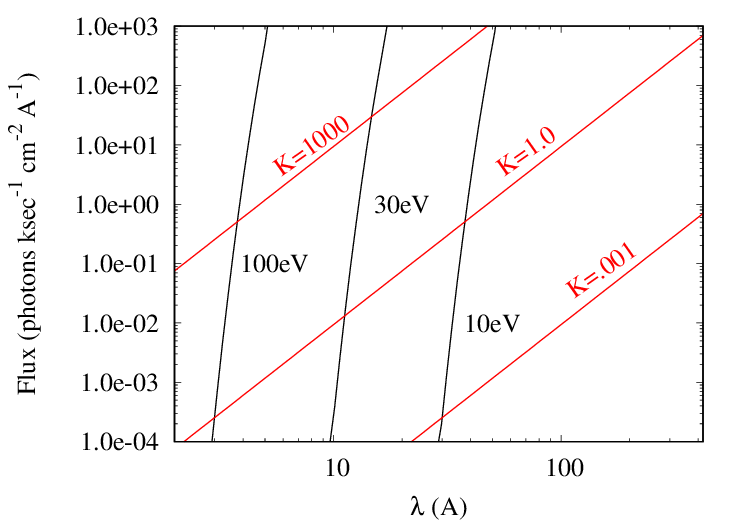}
 \caption{Comparison between TR (red) and UV/X-ray
 emission (black) during the supersoft phase. 
 Curves are labelled by value of $K$ (see text for
 explanation of this parameter) and black body
 temperature of the supersoft source, 
 expressed in eV.\label{compare}}
\end{figure}

In Fig.~\ref{compare}, we compare the magnitude of the
TR and supersoft emission, approximated by a black body
with temperature $T_*$. We take the estimate for the TR
flux in Equation~(\ref{f-mJy1}); the curve corresponding
to the nova parameters that led to this estimate is 
that labelled ``K=1.0'' in Fig.~\ref{compare}. The two
curves labelled ``K=1000'' and ``K=0.001'' allow
for the fact that this may go three orders of magnitude
either way. It seems that even during the supersoft
phase, TR -- if present -- might be significant at
the shortest wavelengths. It might also be the case
that, when the supersoft phase terminates, TR could
be the dominant source of hard radiation.

While the effect of TR may
be significantly lower than our estimate in some cases,
for example, because $V_{\rm com}/{V_e}\ll1$,
or because the dust grains are too small, this may be 
mitigated by the quenching of radiation from the stellar
remnant by the dust. 

If such is the case, TR is likely to have significant
effects on the nova environment, by affecting the 
ionisation balance of the gaseous ejecta, and on 
grain heating. Indeed, where the dust shell is 
optically thick, the effect of TR within the dust
shell may exceed that of radiation from the stellar
remnant, whose radiation can not penetrate the dust.
For example, we speculate whether TR may in part be
responsible for the ``isothermal'' dust phase in 
novae, in which the dust temperature counterintuitively
increases with time after 
$\sim50-100$~days rather than decreases, as would be
expected for dust moving away from a stellar remnant 
of constant bolometric luminosity
\citep{bode83}. \cite{mitchell84} interpreted this behaviour
in terms of grain destruction by chemi-sputtering of carbon
dust, but the production of TR may also play a role in
re-heating the  dust.

\section{Concuding remarks}
\label{discuss}

We have examined the likely importance of transition
radiation, emitted when a relativisic charged particle
enters and leaves a dust particle, in nova environments.
While the direct detection of transition radiation from dusty
novae may be marginal at present, it may be detectable
with future facilities. Even if it is not directly
observable, it may play a significant role in the
evolution of nova remnants. 

More generally, transition radiation may be important
in (a)~dusty supernova remnants such as the Crab Nebula
\citep{temim12}, and  (b)~proto-planetary nebulae, such as
V4334~Sgr (Sakurai's Object), which displays non-thermal
radio emission \citep{hajduk24}, and is extremely dusty
\citep{evans22}, with likely large grains
\citep{bowey21,bowey22}.

\section*{Acknowledgements}
I thank Dipankar Banerjee and Laura Chomiuk for their 
encouragement, and for helpful suggestions and comments
on an earlier version of this paper. I also thank
Kim Page for helpful information about the X-ray
emission of novae.

\section*{Data availability}
Not applicable.

\appendix

\section{Testing the assumption that $E_1=0$.}
We justify our assumption that we can take
$E_1=0$ in the integration over $E$
in Section~\ref{detail}. The full derivation
of Equation~(\ref{TR5})
with $E_1\ne0$ leads to an additional term
in $ [L_\nu]_{\rm TR}$, given by
\begin{equation}
P_1(\beta,\delta) \:\: \frac{\pi{e}^2n_0}{[mc^2]^4} \: \frac{M_{\rm g}}{\rho} \: \frac{V_{\rm com}}{V_{\rm g}} \: a_2^{(4-\delta)} 
 \left ( \frac{\omega_{\rm p}}{\omega} \right )^4 \times
   ( {2\omega_{\rm p}mc}  )^{(5-\delta)} \: 
  \left ( \frac{E_1}{E_2}\right )^{5-\delta}   \:\:,
 \label{TR5-A}
\end{equation}
where $P_1(\beta,\delta)$
\begin{equation}
 =\frac{(4-\beta)}{(3-\beta)(1-\beta)}
 \left [ \frac{1-\beta}{5-\delta} - \eta^2 \: 
 \frac{3-\beta}{7-\delta}
 + \frac{2\eta^{3-\beta}}{8-\beta-\delta}\right ] \:\:.
 \label{P1}
\end{equation}
Here $\eta=a_{c1}/a_2$ and 
\[ a_{c1}=  \frac{c}{2\omega_p} \:\: \left ( \frac{E_1}{mc^2}
  \right ) \:. \]
Hence
  \begin{equation}
 [L_\nu]_{\rm TR} \propto \left \{  P_0 - \left ( \frac{E_1}{E_2} 
 \right )^{5-\delta}P_1    \right \} \:\:.
  \end{equation}
As we must have that $\eta=a_{c1}/a_2<1$, and if 
$E_1\ll{E_2}$ as seems reasonable, the assumption
that $E_1=0$ is justified provided that
$\delta<5$, (i.e. $\alpha<2$). This is indeed the case for the
objects we have considered; the assumption that $E_1=0$ can be
therefore be justified.

%%%%%%%%%%%%%%%%%%%%%%%%%%%%%%%%%%%%%%%%%%%%%%%%%%

% \include{bibfile}
% Don't change these lines
\bsp	% typesetting comment
\label{lastpage}
\end{document}